\documentclass{iaus}
\usepackage{graphicx}
\usepackage{natbib}

\title[Evolution of SNe Ibc \& GRB progenitors ] %% give here short title %%
{Evolution of progenitor stars of Type Ibc supernovae and long gamma-ray bursts}

\author[Yoon et al.]   %% give here short author list %%
{S.-C. Yoon$^1$,
 N. Langer$^2$,
 M. Cantiello$^2$,
 S. E. Woosley$^1$,
 \and G. A. Glatzmaier$^3$}

\affiliation{$^1$Department of Astronomy \& Astrophysics, University of California, Santa Cruz \\ High Street,
Santa Cruz, CA 95064, USA  \\[\affilskip]
$^2$ Astronomical Institute, Utrecht University, Utrecht, The Netherlands \\[\affilskip]
$3$ Department of Earth \& Planetary Sciences, University of California, Santa Cruz \\
High Street, Santa Cruz, CA 95064, USA
}

\pubyear{2008}
\volume{250}  %% insert here IAU Symposium No.
\pagerange{1--6}
\jname{Massive Stars as Cosmic Engines}
\editors{F. Bresolin, P.A. Crowther \& J. Puls, eds.}
\begin{document}

\maketitle

\begin{abstract}
We discuss how rotation and binary interactions may be related to the diversity 
of type Ibc supernovae and long gamma-ray bursts. After presenting recent evolutionary 
models of massive single and binary stars including rotation, the Tayler-Spruit dynamo
and binary interactions, we argue that the nature of SNe Ibc progenitors from binary systems
may not significantly differ from that of single star progenitors in terms of rotation, 
and that most long GRB progenitors may be produced via the quasi-chemically homogeneous evolution
at sub-solar metallicity. We also briefly discuss the possible role of magnetic fields
generated in the convective core of a massive star for the transport of angular momentum, 
which is potentially important for future stellar evolution models of supernova and GRB progenitors.
\end{abstract}

\firstsection % if your document starts with a section,
\section{Introduction}

Rotation influences not only the evolution of massive stars, 
but also their supernova (SN) explosions \citep{Maeder00,Heger00}.
In particular, recently many asymmetric supernovae
with unusually large energy (broad-lined SNe or hypernovae) have been discovered
(e.g. \citealt{Mazzali07}),  
which shows evidence for rapidly rotating progenitors. 
The most spectacular example may be long gamma-ray bursts (GRBs), 
which are generally believed to be produced by deaths 
of some massive stars retaining extremely large angular momenta
in their cores ($j \ge \sim 10^{16}~\mathrm{cm^2~s^{-1}}$; \citealt{Woosley93, Macfadyen99}; 
see, however, \citealt{Dessart08}). 
Interestingly, such energetic core-collapse events seem to only occur in
Wolf-Rayet (WR) stars: all of the broad-lined SNe/Hypernovae and the supernovae associated with 
long GRBs have been observationally identified as Type Ic (see \citealt{Woosley06b} for a review). 
This raises the question which WR stars can produce
broad-lined SNe Ic or long GRBs while most WR stars die as normal SNe Ibc. 
Here we present recent evolutionary models of massive stars
that include binary interactions and the transport of chemical species and angular momentum
via rotationally induced hydrodynamic instabilities and magnetic torques,
and discuss how rotation and binary interactions may be related 
to the diversity of SNe Ibc and long GRBs. 

\section{Single star models}

\begin{figure}
\begin{center}
\resizebox{0.40\hsize}{!}{\includegraphics{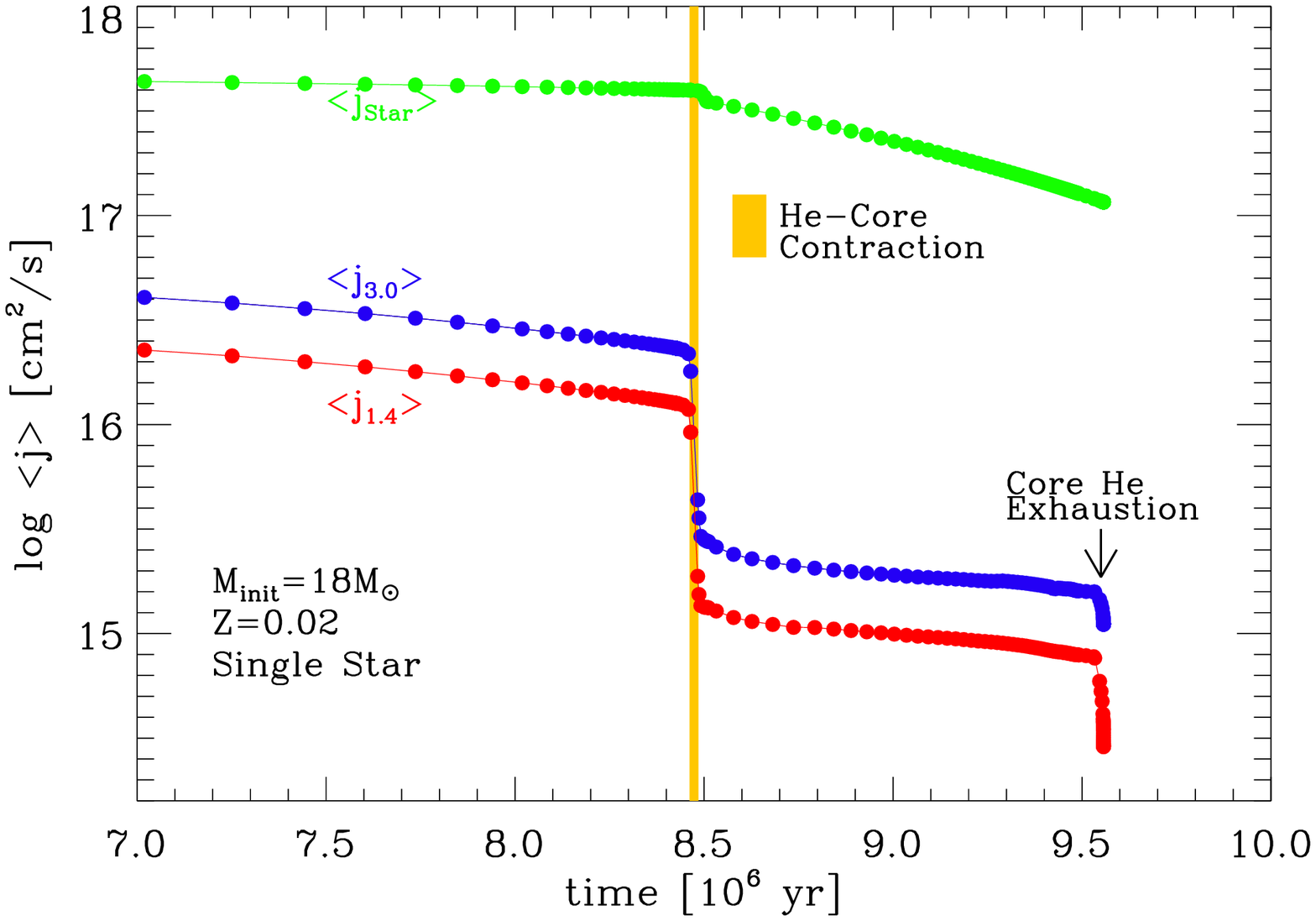}}
\resizebox{0.40\hsize}{!}{\includegraphics{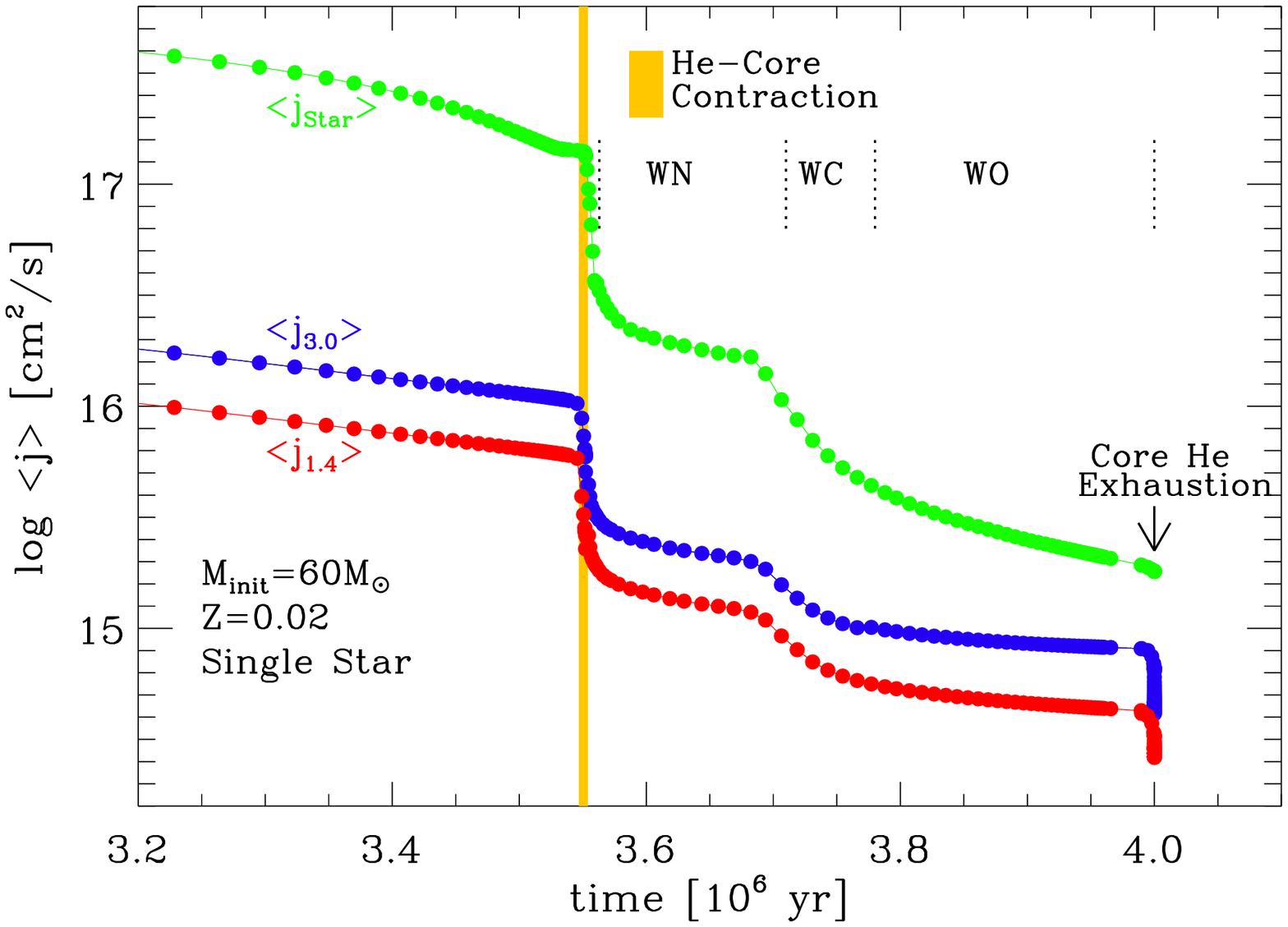}}
\caption{Mean specific angular momentum of the star and the innermost $1.4~\mathrm{M_\odot}$ and $3.0~\mathrm{M_\odot}$ 
as a function of the evolutionary time for $18~\mathrm{M_\odot}$ (left panel) and $60~\mathrm{M_\odot}$ (right panel) 
models.  
The time span for the helium core contraction is marked by the color shade
as indicated by the label.}\label{fig:jcore1}
\end{center}
\end{figure}

Redistribution of angular momentum and chemical species in a rotating star
occurs by rotationally induced hydrodynamic instabilities
and magnetic torques (See \citealt{Talon07} for a review). 
Eddington-Sweet circulations, shear instability and 
Goldreich-Schubert-Fricke instability among others have been considered 
in previous non-magnetic models \citep{Maeder00,Heger00, Hirschi04}, 
and the so-called Tayler-Spruit
dynamo \citep{Spruit02} has been implemented in recent magnetic models (e.g. \citealt{Heger05}; \citealt{Maeder05}; 
\citealt{Yoon05}). 
In non-magnetic models, 
it is shown that the buoyancy due to the chemical gradient
at the interface between the core and the envelope largely prohibits
the considered rotationally induced hydrodynamic instabilities from 
transporting angular momentum. The amount of angular momentum retained
in the core at the pre-supernova stage is thus close to its initial value even
at solar metallicity (\citealt{Heger00}; \citealt{Hirschi04}). 
Most massive stars are predicted to die
with an enough amount of angular momentum in the cores to produce
long GRBs via formation of millisecond magnetars or collapsar 
(i.e., $j\ge \sim 10^{16}~\mathrm{cm^2~s^{-1}}$),
given that a large fraction of young massive stars in our Galaxy and Small/Large Magellanic Clouds
are rapid rotators (e.g. \citealt{Maeder00}; \citealt{Mokiem06}; \citealt{Hunter08}).
On the other hand, in magnetic models adopting the Tayler-Spruit dynamo
the core is effectively spun down by magnetic torques, 
and the predicted spin rates of white dwarfs and young neutron stars
are smaller by two orders of magnitude than those from non-magnetic models,  
which can better explain observations \citep{Heger05, Suijs08}.

Fig.~\ref{fig:jcore1} shows the evolution of the core angular momentum
in the magnetic model sequences with 
$M_\mathrm{init} = 18~\mathrm{M_\odot}$ \& $v_\mathrm{rot,init} = 144~\mathrm{km~s^{-1}}$,   
and $60~\mathrm{M_\odot}$ \& $v_\mathrm{rot,init} = 186~\mathrm{km~s^{-1}}$.
In the sequence with $M_\mathrm{init} = 18~\mathrm{M_\odot}$, 
the core loses a significant amount of angular momentum 
during the helium core contraction phase 
where a strong degree of differential rotation between the core and the envelope 
 appears.
A similar effect is also observed during the CO core contraction phase.
In the sequence with $M_\mathrm{init} = 60~\mathrm{M_\odot}$, 
spinning-down of the core during He-core contraction becomes less significant 
as the star loses the hydrogen envelope, which 
leads to a smaller moment of inertia of the envelope.
However, the core is further spun down by loss of mass due to LBV/WR winds
during core He burning.
At the neon burning stage, both stars retain a similar amount of angular momentum in their cores
as shown in Fig.~\ref{fig:jcore1}. In fact, the calculations by \citet{Yoon06} show that
magnetic models give $<j_\mathrm{1.4}> \simeq 2...3  \times 10^{14}~\mathrm{cm^2~s^{-1}}$ for different
initial metallicities, masses and rotational velocities, implying that
most massive stars including Type Ibc progenitors should die with 
a similar amount of core angular momentum.
This conclusion remains the same even for binary stars, as discussed below. 

\begin{figure*}
\center
\resizebox{0.38\hsize}{!}{\includegraphics{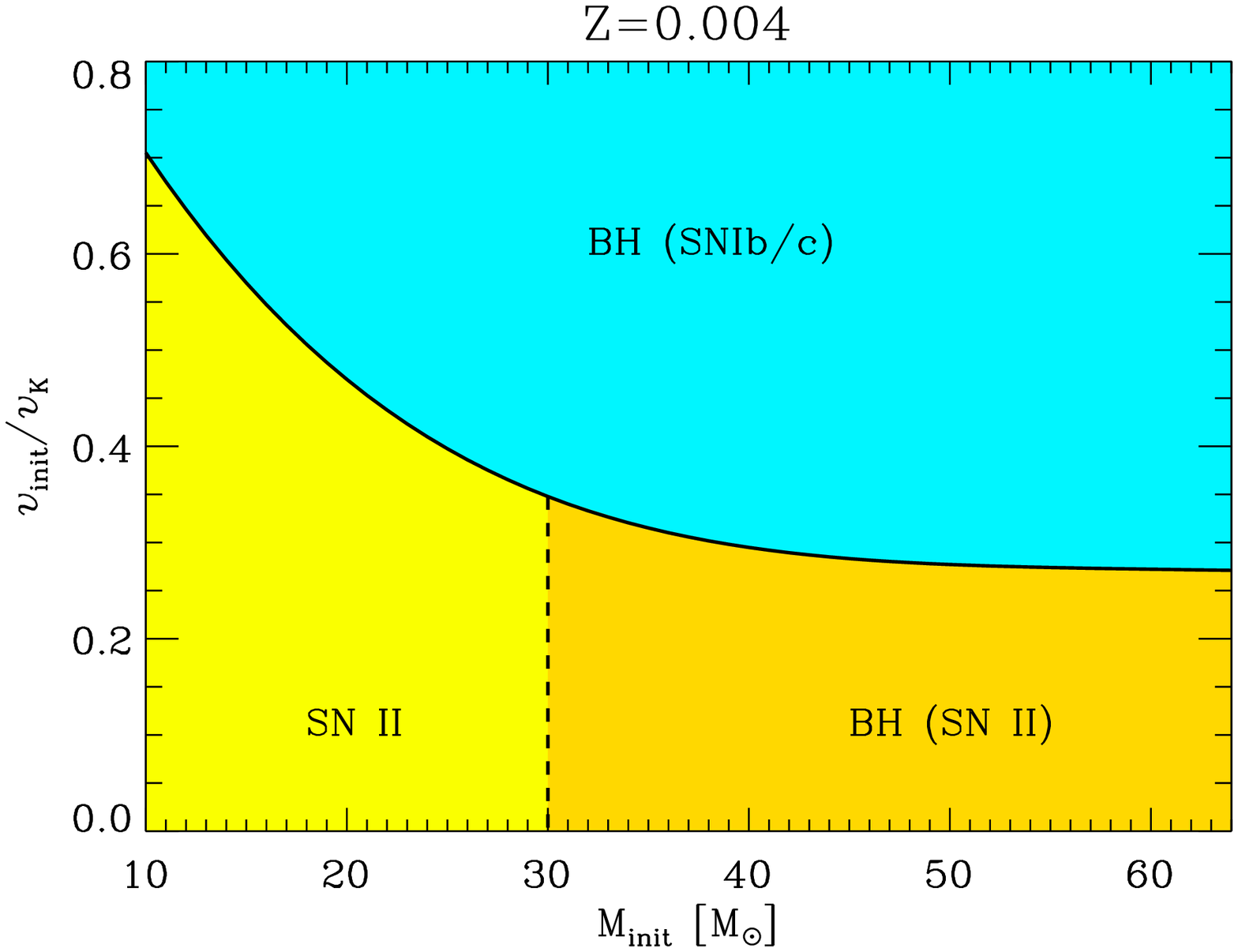}}
\resizebox{0.38\hsize}{!}{\includegraphics{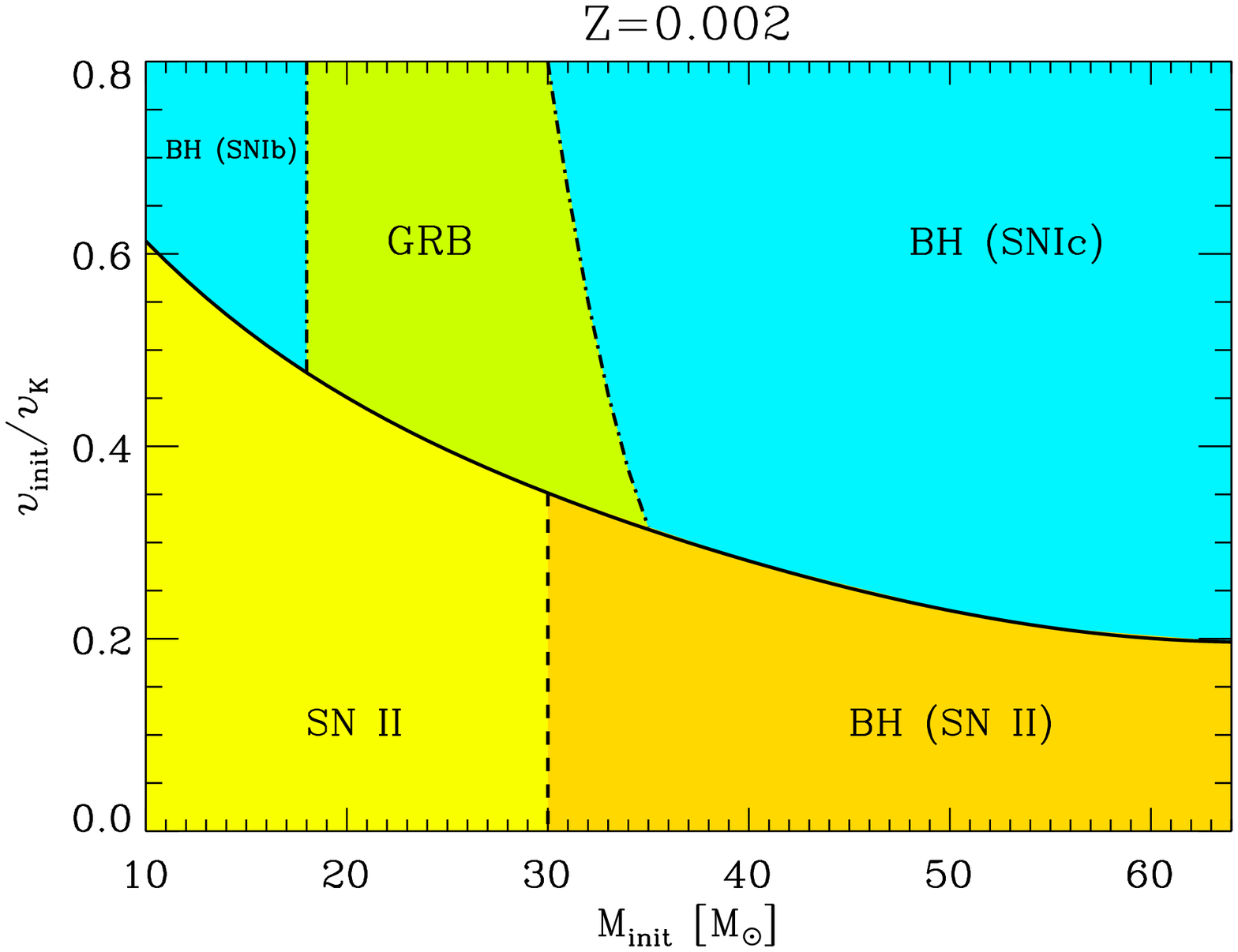}}
\resizebox{0.38\hsize}{!}{\includegraphics{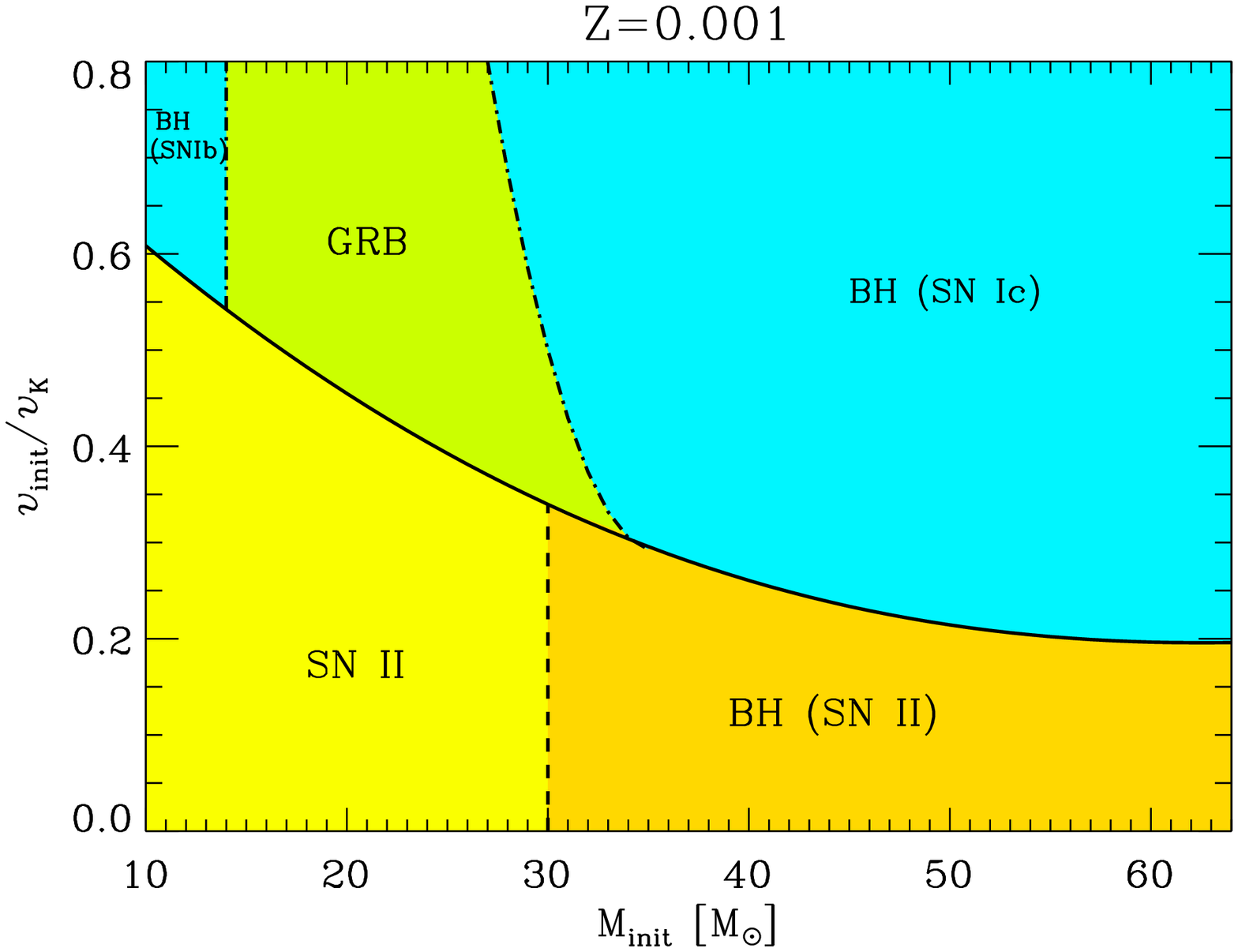}}
\resizebox{0.38\hsize}{!}{\includegraphics{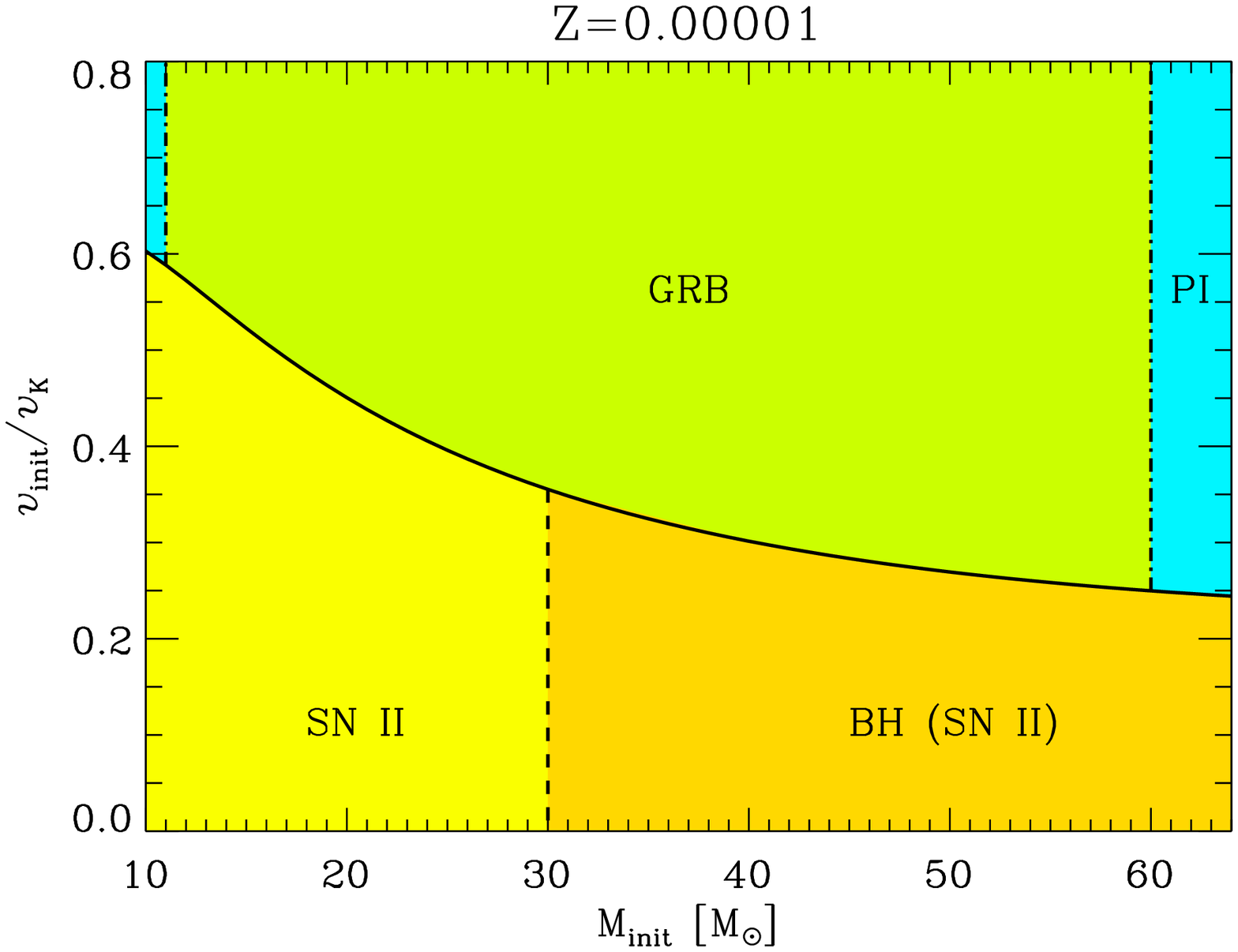}}
\caption{Final fate of our rotating massive star models at four different metallicities ($Z=$ 0.004, 0.002, 0.001, 
\& 0.00001), in the plane of initial mass and 
initial fraction of the Keplerian value of the equatorial rotational velocity.
The solid line divides the plane into two parts, where stars evolve quasi-chemically homogeneous
above the line, while they evolve into the classical core-envelope structure below the line.
The dotted-dashed lines bracket the region of quasi-homogeneous evolution 
where the core mass, core spin and stellar radius are compatible with the collapsar 
model for GRB production (absent at Z=0.004). 
To both sides of the GRB production region for $Z$ = 0.002 and 0.001, black holes are 
expected to form inside WR stars, but the core spin is insufficient to allow GRB production.  
For $Z$ = 0.00001, the pair-instability might occur to the right side of the 
GRB production region, although the rapid rotation may shift
the pair instability region to larger masses.
The dashed line in the region of non-homogeneous evolution 
separates Type II supernovae (SN II; left) and black hole (BH; right) formation,
where the minimum mass for BH formation is simply assumed to be $30~\mathrm{M_\odot}$. 
From \citet{Yoon06}. 
}\label{fig:ches2}
\end{figure*}

An exception is the case for the so-called chemically homogeneous evolution. 
If the initial rotational velocity is exceptionally high and if metallicity 
is sufficiently low, chemical mixing by Eddington-Sweet circulations
may occur on a time scale even smaller than the nuclear time scale. 
Quasi-homogeneity of the chemical composition of a star is thus ensured on the main sequence
and the star is gradually transformed into a massive
WR star, avoiding the giant phase that would result in 
strong braking down of the core. 
All of the necessary conditions for producing long GRBs  -- massive core to make
a black hole, removal of hydrogen envelope and 
retention of a large amount of angular momentum
in the core  -- thus can be fulfilled by this type of evolution
\citep{Yoon05, Woosley06}.  
This chemically homogeneous evolution scenario (CHES) 
favors low metallicity environment for producing long GRBs 
(Fig.~\ref{fig:ches2})
and predicts a higher ratio of GRB to SN rate at higher redshift, 
which should be tested by future observations (\citealt{Yoon06}; cf. \citealt{Kistler07}).

\section{Binary star models}

\begin{figure}
\begin{center}
\resizebox{0.45\hsize}{!}{\includegraphics{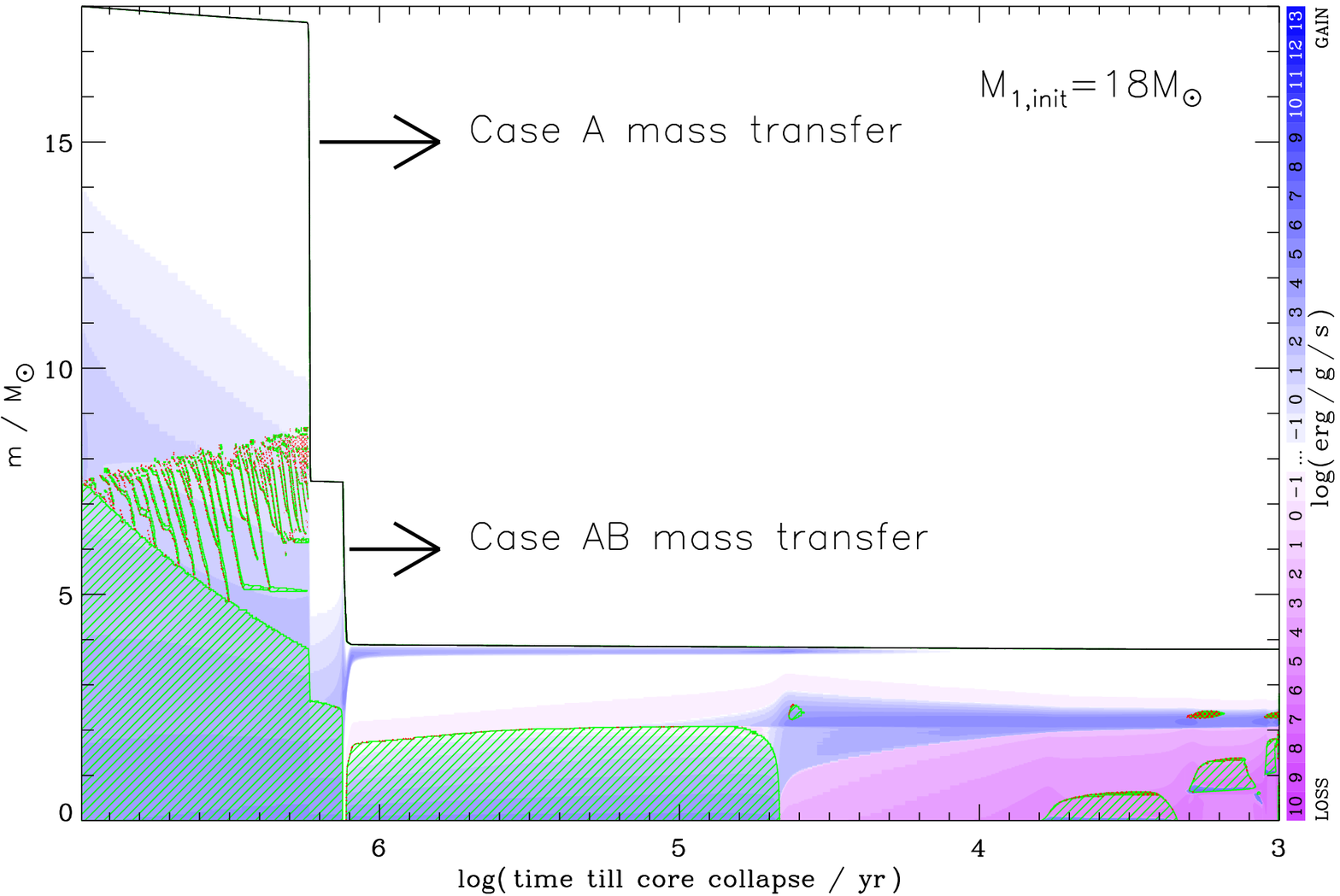}}
\resizebox{0.45\hsize}{!}{\includegraphics{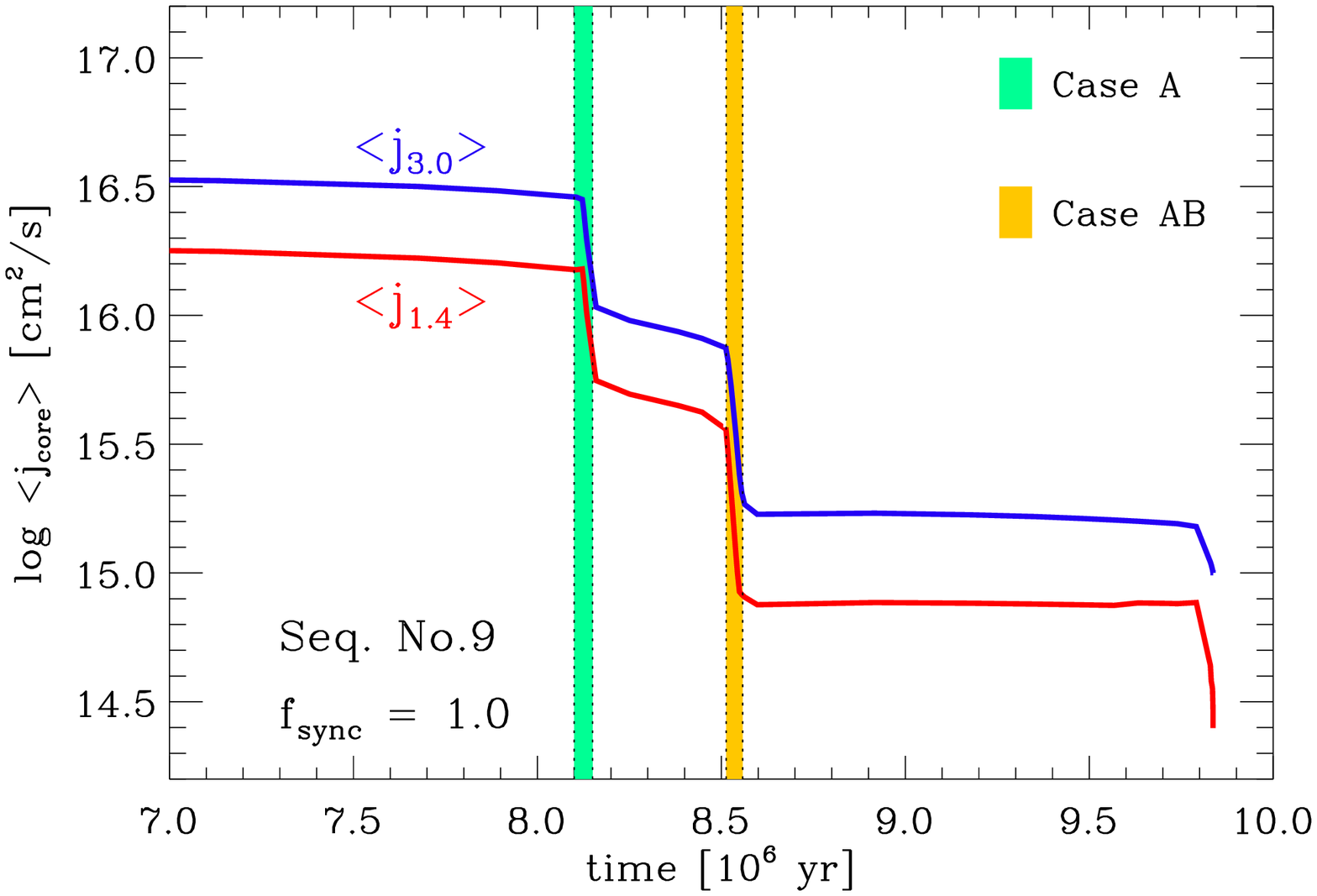}}
\caption{\textit{Left panel} Evolution of the internal structure of the primary 
star in a binary system of $M_\mathrm{primary, init} = 18~\mathrm{M_\odot}$,  
$M_\mathrm{secondary, init} = 17~\mathrm{M_\odot}$ and $P_\mathrm{init} = 4~\mathrm{days}$, 
from zero age main sequence to neon burning.
\textit{Right panel} Mean specific angular momentum of the innermost $1.4~\mathrm{M_\odot}$ and $3.0~\mathrm{M_\odot}$ 
of the primary star considered in the left panel as a function of time.
The time span for Case A or Case AB mass transfer phase is marked by the color shades
as indicated by the labels.}\label{fig:bjcore}
\end{center}
\end{figure}

A significant fraction of SNe Ibc may be produced in close binary systems
(e.g. \citealt{Podsiadlowski92}). 
An example is given in Fig.~\ref{fig:bjcore} that shows the evolution of 
of the primary star in a close binary system with $P_\mathrm{init}=4~\mathrm{days}$, 
$M_\mathrm{primary,init} = 18~\mathrm{M_\odot}$, and
$M_\mathrm{secondary,init} =17~\mathrm{M_\odot}$. 
Both stars are tidally synchronized early on the main sequence. 
Once the primary star fills the Roche-lobe radius, it  loses about 
$7~\mathrm{M_\odot}$ during the Case A mass transfer phase, and additional $3.5~\mathrm{M_\odot}$
later during the Case AB mass transfer phase, becoming a $4~\mathrm{M_\odot}$ WR star. 
The core loses angular momentum mostly during these Case A and AB mass transfer phases
as shown in Fig.~\ref{fig:bjcore}. The amount of angular momentum in the core
at the neon burning stage turns out to be very similar to those in single star models 
(see Fig.~\ref{fig:jcore1}). 
In fact, we find that 
mean specific angular momentum
of the innermost $1.4~\mathrm{M_\odot}$ at the final evolutionary stage of a primary star in a binary system
does not change much according to different initial parameters (primary mass, mass ratio and orbital separation): it
remains within a narrow range of $2...3.5 \times 10^{14}~\mathrm{cm^2~s^{-1}}$ regardless
of the detailed history of binary interactions, as long as the tidal synchronization
is not unusually strong (Yoon, Woolsey \& Langer 2008, in prep.). 
This implies that the nature of most binary star progenitors of SNe Ibc may not much differ from 
that of single star progenitors, in terms of rotation.

The evolution of secondary stars has not yet been well understood. 
In particular, it sensitively depends on the uncertain efficiency of semi-convection 
whether the mass accreting star may be rejuvenated or not \citep{Braun95}. 
If a rather large semi-convection parameter is adopted, 
rejuvenation can significantly weaken the chemical gradient
between the hydrogen burning core and the envelope, 
in favor of rotationally induced chemical mixing.
As the secondary is spun up to the critical rotation by mass accretion, 
even the chemically homogeneous evolution can be occasionally induced 
if metallicity is sufficiently low and 
if the secondary is not strongly spun down by the tidal synchronization after the mass
accretion phase \citep{Cantiello07}. The secondary will  eventually die as a GRB after traveling from a few 
to several hundreds PCs away, if the binary system is unbound due to the supernova kick
as a result of the explosion of the primary.
This scenario may explain the recent observational evidence that
some GRBs are produced in runaway stars \citep{Hammer06}. 

Other types of binary interactions may also lead to formation of rapidly rotating WR stars 
to produce long GRBs. Tidal spinning-up of a WR star in a compact binary system with 
a neutron star or a black hole companion \citep{Brown00,Izzard04, vanPutten04, vandenHeuvel07}  
and merger of two helium cores in 
a common evenlope \citep{Fryer05} have been recently suggested among others. It remains uncertain, however, 
that such binary systems could explain the observed GRB rate. For example, recent stellar
evolution models by \citet{Detmers08} show
that a merger of the WR star with the compact object, which is not supposed to produce
a classical long GRB, is the most likely outcome in the former case.
Further detailed evolutionary models are certainly needed 
for observationally testing different evolutionary scenarios (e.g. \citealt{vanMarle08}). 
On the other hand, 
it is puzzling why no GRB-associated SNe Ib have been observed yet
while most GRB progenitor scenarios predict their existence (e.g. \citealt{Yoon06}).
Within the CHES, this puzzle might be solved
if one considered anisotropic mass loss, as discussed in \citet{Meynet07}.  

\section{Concluding remarks}

\begin{figure}
\begin{center}
\resizebox{0.85\hsize}{!}{\includegraphics{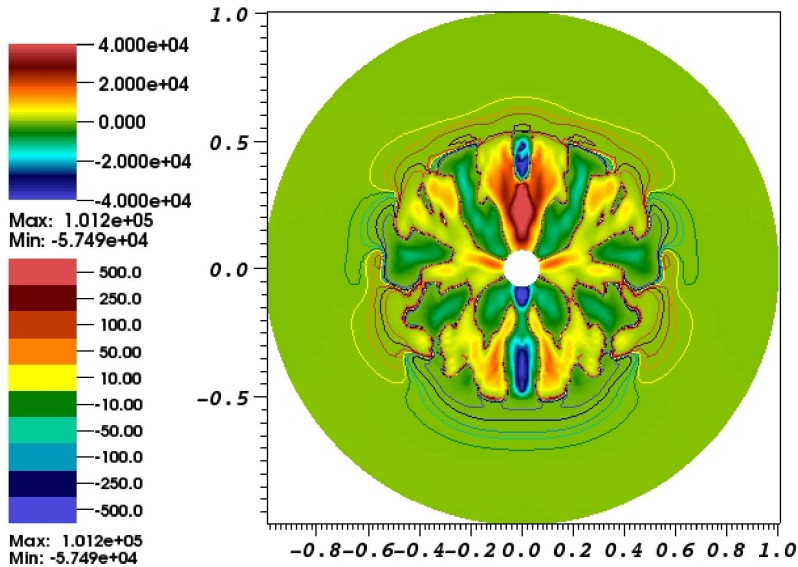}}
\caption{Mean radial fields $B_\mathrm{r} (r, \theta)$ on the meridional plane
in a $12~\mathrm{M_\odot}$ rotating star on the main sequence
in a MHD simulation with a 3-D anelastic code \citep{Glatzmaier84}. 
The adopted angular velocity is $10^{-5}~\mathrm{Rad~s^{-1}}$. 
The inner region of $r \le 7 \times 10^{10}~\mathrm{cm}$ ($r \le 0.5$ in the code units) 
is the convective core. 
}\label{fig:bfields}
\end{center}
\end{figure}

Our stellar evolution models including the Tayler-Spruit dynamo indicate that 
most SNe Ibc progenitors should explode with a similar amount of angular momentum
in their cores, regardless of their single or binary star origin. 
This is due to the self-regulationary nature of the Tayler-Spruit dynamo.
Loss of the hydrogen envelope due to stellar winds or mass transfer results in 
both removal of angular momentum from the core 
and weakening of the core braking by the extended hydrogen enveloped 
due to magnetic torques, and vice versa. Therefore, most different pre-supernova evolutionary 
paths may not contribute much to the diversity of Type Ibc supernovae in terms of core angular momentum,
although different iron core masses, and thus different spin rates of young neutron stars
may result \citep{Heger05}.   
GRB progenitors, which require unusually large angular momenta, may undergo
the chemically homogeneous evolution, which may not be unusual at low metallicity.
On the other hand, recent observations indicate that not all broad-lined SNe Ic
are associated with long GRBs (e.g. \citealt{Modjaz08}), which needs a theoretical explanation in future work.

Although the predicted spin rates of the stellar 
remnants from the magnetic models are consistent with
observations, the validity of the Tayler-Spruit dynamo has been recently questioned
by several authors \citep{Denissenkov07, Zahn08}. Furthermore, we might have ignored potentially  important physical 
ingredients in simulating the evolution of rotating massive stars. 
These include gravity waves (Townsend in this volume),
and magnetic fields generated by the convective core.
For instance, our recent 3-D simulations with an anelastic magnetohydrodynamics code \citep{Glatzmaier84}
show that the strength of poloidal fields generated in the convective core in a young massive
star may amount to several thousand Gauss on average (Fig.~\ref{fig:bfields}), and its influence on the transport
processes might be comparable to what the Tayler-Spruit dynamo predicts.
This issue will be addressed in Yoon, Woosely \& Glatzmaier (2008, in prep.).

This work was, in part, supported by the DOE Program 
for Scientific Discovery through Advanced Computing
and NASA.

\end{document}